# Unraveling the interface chemistry between HCN and cosmic silicates by the interplay of infrared spectroscopy and quantum chemical modeling

Niccolò Bancone ,[†,‡] Rosangela Santalucia ,[‡] Stefano Pantaleone ,[‡] Piero Ugliengo ,[‡] Lorenzo Mino ,[‡] Albert Rimola ,[*,†] and Marta Corno [*,‡]

†*Departament de Química, Universitat Autònoma de Barcelona, Bellaterra, 08193, Catalonia, Spain*

‡*Dipartimento di Chimica and Nanostructured Interfaces and Surfaces (NIS) Centre, Università degli Studi di Torino, via P. Giuria 7, 10125, Torino, Italy.*

E-mail: albert.rimola@uab.cat; marta.corno@unito.it

Phone: +34-935813723; +39-0116702439

## Abstract

Understanding the interaction between hydrogen cyanide (HCN) and silicate surfaces is crucial for elucidating the prebiotic processes occurring on interstellar grain cores, as well as in cometary and meteoritic matrices. In this study, we characterized the adsorption features of HCN on crystalline forsterite ($Mg_2SiO_4$) surfaces, one of the most abundant cosmic silicates, by combining experimental infrared spectra at low temperatures (100–150 K) with periodic DFT simulations. Results showed the coexistence of both molecular and dissociative HCN adsorption complexes as a function of the considered forsterite crystalline face. Molecular adsorptions dominate on the most stable surfaces, while dissociative adsorptions occur predominantly on surfaces of lower stability, catalyzed by the enhanced Lewis acid-base behavior of surface-exposed $Mg^{2+}$–$O^{2-}$ ion pairs. On the whole set of adsorption cases, harmonic frequency calculations were carried out and compared with the experimental infrared bands. To disentangle each vibrational mode contributing to the experimental broad bands, we run a best non-linear fit between the predicted set of frequencies and the experimental bands. The outcome of this procedure allowed us to: i) deconvolute the experimental IR spectrum by assigning computed normal modes of vibrations to the main features of each band; ii) reveal which crystal faces are responsible of the largest contribution to the adsorbate vibrational bands, giving information about the morphology of the samples. The present straigthforward procedure is quite general and of broad interest in the fine characterization of the infrared spectra of adsorbates on complex inorganic material surfaces.

# Introduction

Hydrogen cyanide (HCN) is recognized as an ubiquitous molecule in various astrophysical environments, ranging from diffuse and dense clouds,[1,2] protostellar hot cores,[3–5] cometary comas[6–8] and meteorites,[9] to planetary atmospheres,[10,11] including Pluto and Saturn's moon Titan.[12–14]

Several studies have highlighted the importance of HCN in the synthesis of organic compounds relevant to the emergence of life, prompting experimental investigations into its reactivity under different conditions mimicking both interstellar and prebiotic environments,[15–19] which more recently have been extended to a theoretical side.[20–24] Within this context, the significance of HCN lays in its ability to polymerize, leading to the formation of complex organic molecules, among which the adenine nucleobase, a fundamental biomolecular building block for life.[25] The gas-phase polymerization involving only HCN has been reported to be hindered by high kinetic barriers, hard to overcome at the interstellar conditions.[26]

Explorations on HCN and HNC reactive pathways have demonstrated that the introduction of basic or acidic catalysts, and/or other molecules, leads to more feasible mechanisms.[24,27] In liquid-phase, pathways related to HCN polymerization seem to be more promising,[23] due to the presence of basic catalysts.[15,20] Among the few studies on the chemistry of HCN in the solid phase, water ices have been investigated as possible concentrator[28,29] of gaseous HCN and as catalysts for its dimerization.[30] On the other hand, the effectiveness of surface siloxyl radicals (SiO·) as initiators for the polycyclization of HCN has been theoretically described,[31] thus introducing the potential role of silicate materials as heterogeneous catalysts for HCN prebiotic reactions.

Silicates represent the most abundant minerals in the Universe as they are the major refractory material in interstellar grains, cometary nuclei, meteoritic chondrites and rocky planets; as an emblematic example, they represent the most abundant and important inorganic material in Earth's crust. Silicates, moreover, possess well-established catalytic properties towards chemical reactions of interstellar[32,33] and prebiotic interest.[34] Therefore, investigating the activation of HCN upon silicate adsorption and the subsequent catalytic reactivity (particularly in the absence of external energy inputs) is of interest in the astrochemistry, cosmochemistry and prebiotic chemistry fields.

In a previous work by some of us, spectroscopic experimental measurements of HCN in interaction with synthetic Mg-rich silicate surfaces (both amorphous and crystalline) were performed, with the aim of investigating their catalytic properties in the HCN polymerization.[36] The chosen temperature range (i.e., 100-300 K) aligned well with the conditions found in common prebiotic environments, including interplanetary regions, meteorites, and comets. The study employed innovative methods for the safe production and controlled dosing of pure HCN onto solid samples, followed by Fourier-transformed infrared (FT-IR) spectroscopy and high-resolution mass spectrometry (HR-MS) analysis to characterize the reaction products. For what concerns the adsorption of gaseous HCN, results showed that above 150 K the HCN/silicates complexes undergo an intricate chemistry mainly due to the presence of surface-exposed $Mg^{2+}$–$O^{2-}$ ion pairs, as they present an enhanced Lewis acid-base behavior[37] that enable the deprotonation of adsorbed HCN, which in turn triggers their polymerization. However, the first steps of the HCN adsorption process were not investigated, and accordingly the study lacked a detailed and a thorough atomistic description of the molecule-surface interactions.

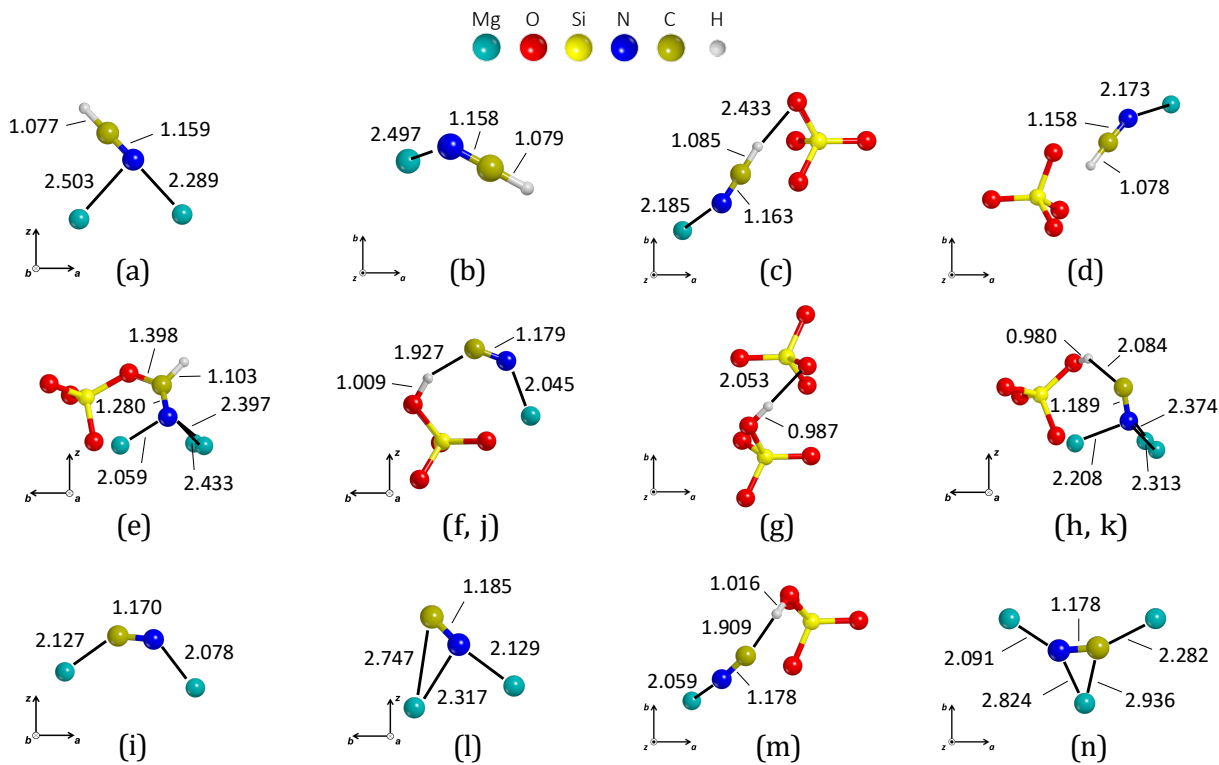

Figure 1: Representations of the optimized geometries (focusing on the binding region). All distances are in Å. Adapted with permission from 35. Copyright 2023 Royal Society of Chemistry.

Due to that, a recent computational work[35] on the HCN adsorbed on $Mg_2SiO_4$ (forsterite) surfaces was published, being dedicated to characterize the structures and energetics of all possible HCN adsorption complexes (see Figure 1), while vibrational frequencies were only marginally discussed. Indeed, the lack of an experimental counterpart made impossible the assignment of the contribution of each adsorption mode. At the same time, the experimental spectra were rather intricate, with probably overlapping bands difficult to be assigned.

The present work aims at overcoming the above-mentioned drawbacks, by combining in a more thorough way the experimental and theoretically predicted infrared spectra involving the adsorbate modes. The main outcome allows a fine interpretation of the experimental bands, providing a band deconvolution as least biased as possible, and also some hints about the crystal morphology of the forsterite samples and its affectation in the HCN adsorption.

# Methods

## Materials

Crystalline forsterite was obtained by heating the pristine amorphous magnesium silicate at 1073 K for 24 h in air. The specific surface area was of $\approx$ 26 m$^2$/g. Before HCN dosage, the samples were outgassed at 673 K for 1 hour under high vacuum (residual pressure 5×10$^{-4}$ mbar) to obtain a dehydrated/dehydroxylated surface. More information about the properties and the synthesis of the two materials are available in ref. 37. The HCN used in the experiment was synthesized using a safe and efficient procedure developed by our group, which is detailed in the supporting information of ref. 36.

## Experimental methodology

For the transmission IR spectroscopic measurements, the powder sample was pressed in the form of self-supporting pellet and placed into a quartz cell equipped with IR-transparent KBr windows and a valve for connection to a vacuum line (residual pressure 5×10$^{-4}$ mbar). This setup allows *in situ* adsorption/desorption experiments to be conducted under controlled atmosphere. Additionally, the cell is designed for low-temperature experiments (approximately 100 K) by cooling the sample with a reservoir of liquid nitrogen. In the experiment, HCN (5 mbar equilibrium pressure) was dosed into the cooled cell. Under these conditions, the gas condensed on the walls of the cell. Subsequently, the system temperature was allowed to increase freely, causing HCN to be slowly released and reach the sample as its temperature approached $\approx$150 K. The initial spectra sequence, discussed in this study, can be essentially described as the dosing of increasing amounts of HCN at a constant temperature (150 K). The HCN adsorption was monitored *in situ* by continuously acquiring IR spectra using a Bruker EQUINOX 55 FTIR spectrometer equipped with a DTGS detector. Typically, 128 interferograms (at a resolution of 2 cm$^{-1}$) were acquired for each spectrum to ensure a good signal-to-noise ratio. Further details are reported in ref. 36.

## Computational methodology

All geometry optimizations and vibrational frequency computations were performed with the CRYSTAL17 periodic quantum mechanical code.[38] The DFT-PBE level of theory was used, complemented with the revised version of the Grimme's D2 empirical term (namely, the D*N correction, specifically customized for periodic systems containing Mg$^{2+}$ cations)[39–41] to account for dispersion interactions. The Ahlrichs-VTZ[42] basis set augmented with polarization functions for describing the atoms of HCN and the smaller basis set proposed by Bruno *et al.*[43] for forsterite atoms (8-511G* For Mg, 8-6311G* for Si, and 8-411G* for O) were used. The same methodology as that of ref. 35 was used, so that we strongly refer to that work for further computational details.

Vibrational frequencies were computed numerically within the harmonic approximation and at the Γ point, thus including vibrations in the unit cell, only. Each Hessian matrix element was calculated through the second derivatives of the potential energy of the stationary points by displacing each atom from its equilibrium geometry along each cartesian coordinate by ±0.003 Å along each cartesian axis (i.e., central difference formula). To speed up the calculations, only the HCN atoms and those belonging to the first layer of the surface models were allowed to displace.

Computed vibrational frequencies suffer from a systematic error intrinsic to the DFT method and to the harmonic approximation. As the main purpose of this work is to compare experimental spectra with the computed ones, each HCN computed frequency was corrected by two scaling factors *s*, one for the C–H and another for the C≡N stretching vibrations, by comparing the gas-phase HCN simulated and experimental frequencies. This allowed us to rescale the whole gas phase HCN spectrum by considering that the systematic errors are normal mode dependent, allowing a more sensible and direct comparison with the experimental spectra. Therefore, the following scaling factors were obtained:

$$S_{CH} = \frac{\nu_{exp}}{\nu_{comp}} = \frac{3311}{3370} = 0.9825$$

$$S_{CN} = \frac{\nu_{exp}}{\nu_{comp}} = \frac{2097}{2127} = 0.9859$$

Additionally, since in some adsorption complexes HCN dissociation takes place, silanol (SiOH) surface groups are generated. For the OH stretching vibration of these surface SiOH groups, the scaling factor was obtained as the ratio between the experimental frequency of surface isolated silanols and the corresponding computed ones for those surfaces models exhibiting isolated SiOH groups, that is:

$$S_{OH} = \frac{\nu_{exp}}{\nu_{comp}} = \frac{3744}{3753} = 0.9976$$

## Simulation of a global IR spectrum

In our previous computational work,[35] we identified 16 adsorption complexes considering 6 different crystalline forsterite surfaces. Thus, in order to simulate a single global spectrum taking into account the contribution of each adsorption complex $j$, we first associated each computed frequency $\nu_i$ with a Lorentzian function $L_i^j(\nu)$, defined as:

$$L_i^j(\nu) = I_i^j \Gamma_i^j / [\pi(\nu-\nu_i)^2 + (\Gamma_i^j/2)^2]$$

where $I_i^j$ is the PBE-D*N computed infrared intensity of the stretching mode corresponding to the $\nu_i$ frequency and $\Gamma_i^j$ is the full width at half maximum (FWHM) of the peak.

The experimental IR signals undergo a broadening in the 2500–3700 cm$^{-1}$ zone, due

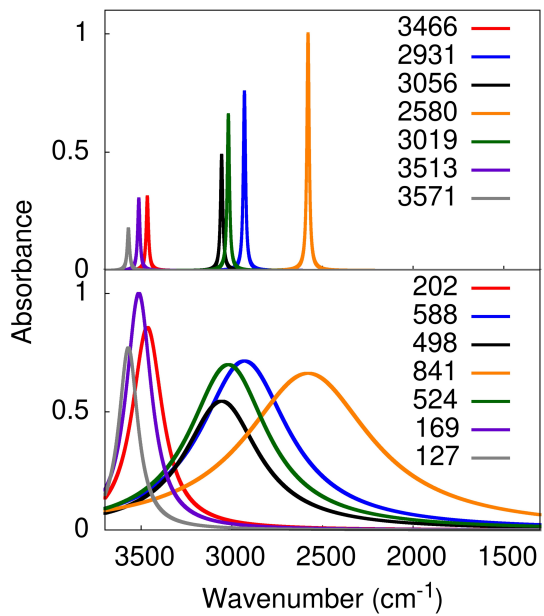

Figure 2: Example of the effect of the Huggins-Pimentel correction[44] on the band-width of the computed O–H stretching bands. Top: simulated bands with a fixed FWHM = 15 cm$^{-1}$. For each signal, the corresponding frequency in cm$^{-1}$ is also reported. Bottom: peaks computed with a FWHM given by the Huggins-Pimentel formula, as adopted in this work. For each signal, the corresponding FWHM in cm$^{-1}$ is also reported.

to the H-bond interactions occurring between the exposed silanols and either the surface-exposed $O^{2-}$ anions or the $CN^-$ adsorbates. In order to better reproduce this broadening of the peaks, we computed $\Gamma_i^j$ for the O–H groups through the formula proposed by Huggins and Pimentel for the correction of the bandwidth as a function of the OH frequency shift with respect to the free SiOH group[44] (see Figure 2):

In order to better reproduce this broadening of the peaks, we computed $\Gamma_i^j$ for the O–H groups through the empirical formula, parametrized on OH and NH H-bonded systems, proposed by Huggins and Pimentel for the correction of the bandwidth as a function of the OH frequency shift with respect to the free SiOH group[44] (see Figure 2):

$$\Gamma_i^j = 0.72\, \Delta_i + 2.5\ cm^{-1}$$

where $\Delta_i$ represents the bathochromic shift of the stretching frequency of the $i$-silanol with respect to a free silanol, so that $\Delta_i$ = 3744 cm$^{-1}$ − $\nu_i$. For the peaks corresponding to the C–H and C≡N stretchings, we instead adopted a fixed $\Gamma^j_i$ = 15 cm$^{-1}$. Accordingly, we computed the IR spectrum of each single $j$-adsorption complex as a sum of the Lorentzian functions associated with its computed signals:

$$S_j(\nu) = \sum_i L^j_i(\nu)$$

and generate the total spectrum $T(\nu)$ through a linear combination of each $S_j(\nu)$:

$$T(\nu) = \sum_j c_j S_j(\nu)$$

where $c_j$ are the weight coefficients of each spectrum, normalized for the case with the best match between experiment and theory.

When HCN deprotonates upon adsorption, the proton is transferred to the surface forming a SiOH group. The remaining CN$^-$ anion interacts with multiple surface-exposed Mg$^{2+}$ cations, forming different CN$^-$–Mg$^{2+}$ interactions of variable strength.[35] Simulations show that the freshly formed SiOH can engage H-bond interactions either with the CN$^-$ or nearby surface-exposed O$^{2-}$ anions, thus experiencing different degrees of perturbation.[35,36] For these reasons, in this work, we treated the C≡N and O–H stretching vibrations separately. That is, we defined two independent $S_j(\nu)$ spectra for C≡N and O–H.

By proceeding this way, we generated a global spectrum $T(\nu)$ by linearly combining 23 single $S_j(\nu)$ spectra, which include: 8 spectra given by the sum of C–H and C≡N signals (molecular adsorptions), 8 spectra given by the C≡N signals alone (dissociative adsorptions) and 7 spectra by the O–H signals (dissociative adsorptions); the remaining one represents the free silanol used to compute the OH stretching scale factor. To match the experimental spectrum, we performed a non-linear best fit between $T(\nu)$ and the experimental one in the 3700–1300 cm$^{-1}$ range, thus excluding any contribution given by the vibrations of the surface. The best fit is performed by changing the relative weights $c_j$ of each spectrum $S_j$. Therefore, only the relative intensities are changed, while keeping the frequencies of the bands fixed at the scaled computed values. From now on, we will refer to the fitted simulated spectrum as $T^{fit}(v)$. Therefore, the value of each $S_j(v)$ contribution in $T^{fit}(v)$ weighs the relevance of each type of HCN/forsterite adsorption complex contributing to the description of the experimental spectrum. In addition, this approach allows us to gauge the contribution of each forsterite crystal face in contributing to the final spectrum, disentangling the experimental bands, as discussed in the next section.

## Results and Discussion

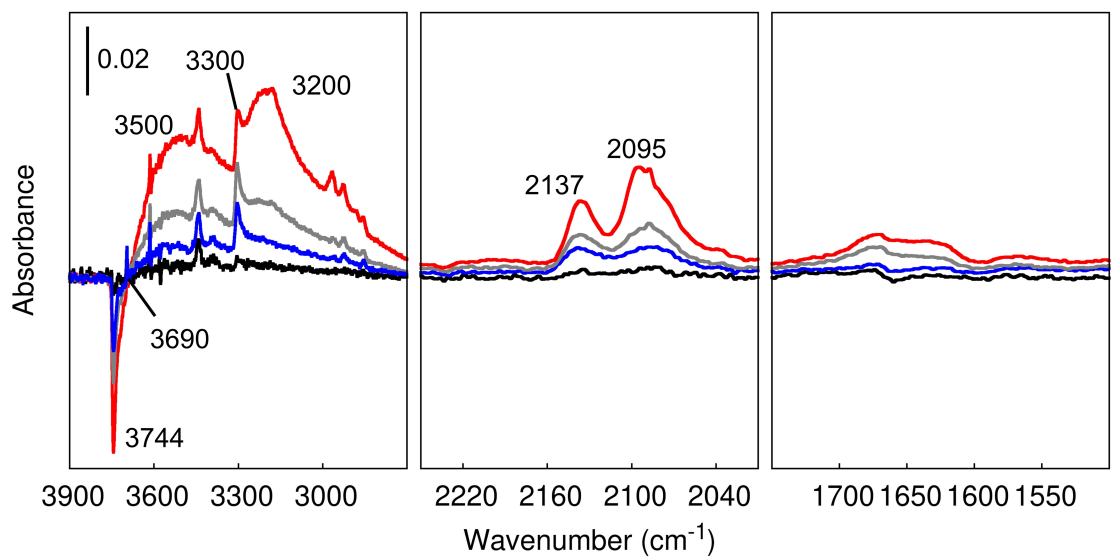

Figure 3: Experimental IR spectra of HCN gas adsorbed at 150 K on crystalline Mg–silicate (forsterite) previously outgassed at 673 K. The series shows the effect of gradually increasing the HCN equilibrium up to 5 mbar (from black to red spectrum). The spectrum of the bare activated material has been subtracted from all spectra. Negative peaks represent species that are consumed after the interaction between HCN and the solid, while positive peaks represent species that are formed.

The experimental IR spectra obtained at 150 K at increasing HCN equilibrium pressure (from black to red) on dehydrated forsterite are reported in Figure 3, while Figure 4 shows the global computed IR spectrum $T^{fit}(v)$ superimposed to the low-coverage experimental one (blue line of Figures 3 and 4), as it better represents the isolated adsorptions modelled in the simulations. Only 14 out of the 23 starting $S_j(v)$ spectra contribute to $T^{fit}(v)$, of which here we report the best 10 cases in Figure 4 and in Table 1. In Table S1 all the adsorption modes are reported with the corresponding vibrational frequencies and weight coefficients to the fitted spectrum.

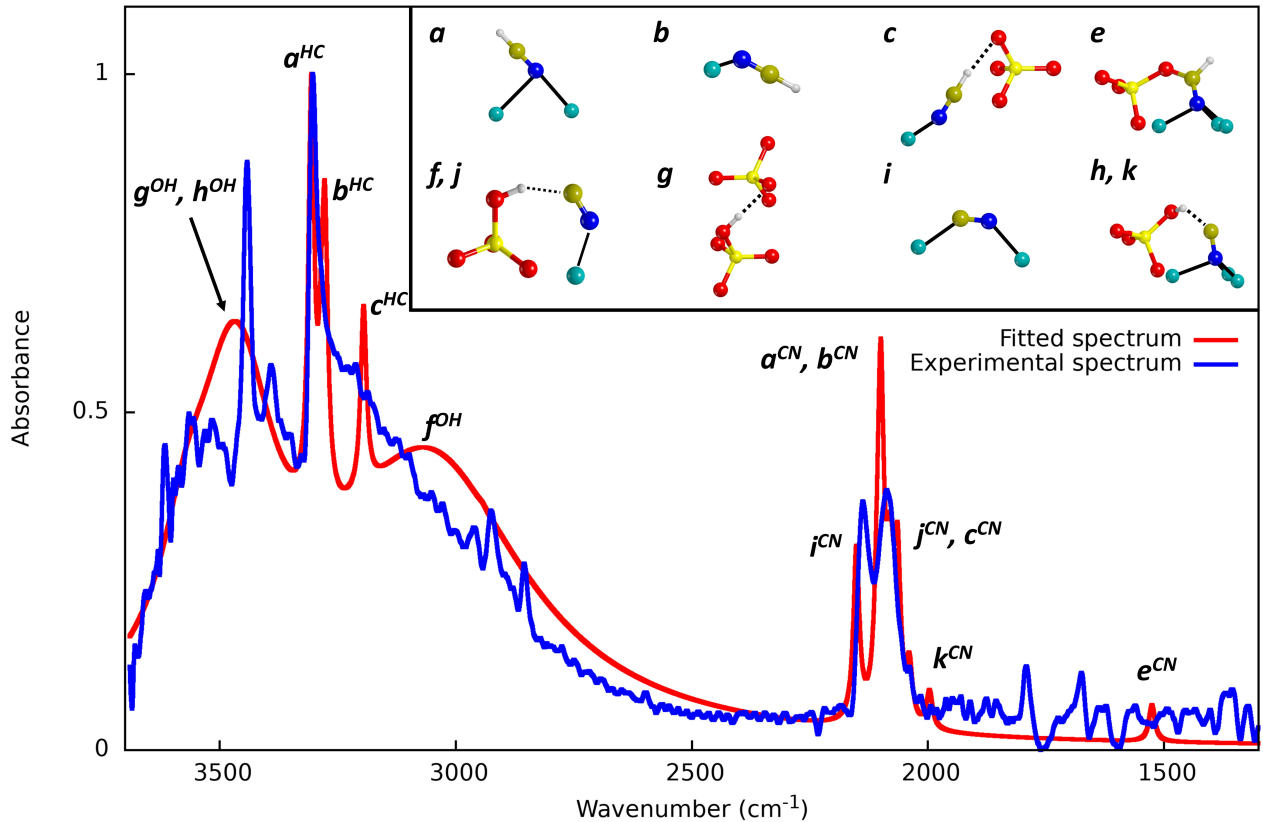

Figure 4: Experimental (blue) and fitted computed (red) IR spectra. The letters correspond to the adsorption complexes which mainly contribute to the experimental spectrum. In the assignments, the corresponding stretching mode, namely HC, CN and OH (silanol), is reported in superscript. Atom color legend: H white, C ochre, N blue, O red, Mg cyan, Si yellow.

Upon dosing HCN, there is a noticeable progressive reduction in the intensity of the band at ca. 3744 (3753 computed) cm$^{-1}$ which represents the isolated silanols, accompanied by the simultaneous growth of two broad absorption bands at ca. 3500 cm$^{-1}$ and 3200 cm$^{-1}$.

This transformation is marked by the appearance of a distinct isosbestic point at ca. 3690 cm$^{-1}$. Previous studies concerning forsterite characterization and the interaction between HCN gas and SiO$_2$ [36] allow for an interpretation of these features, suggesting the formation of SiOH· · ·NCH adducts involving the silanol groups present in the non-completely crystallized portions of the forsterite sample.[37] In our simulations we do not have this specific interaction, as the formation of silanol groups is associated with the HCN deprotonation. Nevertheless, we modelled different degrees of perturbation of the SiOH, thus correctly reproducing the experimental bands in the OH region.

Moreover, according to our simulations, the band at 3200-3300 cm$^{-1}$ is ascribed to $\nu$(C-H) modes (in addition to highly red-shifted $\nu$(O-H)) coupled with the band at 2095 cm$^{-1}$ of the $\nu$(C≡N). This assignments fit well with the computed cases *a* and *b* ($c_i$ = 0.5408 and 0.2173) which belong to the second most stable surface (120) and represent a slightly perturbed molecular HCN, ($\nu$(C-H) = 3305-3276 cm$^{-1}$, $\nu$(C≡N) = 2099 cm$^{-1}$). The *c* case on the (010) surface ($\nu$(C-H) = 3195 cm$^{-1}$, $\nu$(C≡N) = 2064 cm$^{-1}$, $c_i$ = 0.1688) undergoes a larger bathochromic shift of both features with respect to *a* and *b*, because of the stronger interaction of the N atom with a Mg$^{2+}$ cation and the formation of a H-bond with a silicate group Mg$^{2+}$· · ·NCH· · ·OSiO$_3$. This case well explains the C≡N feature below 2095 cm$^{-1}$ while, for the C-H, the presence of a shoulder below 3000 cm$^{-1}$ suggests that many structures similar to *c* could contribute to the observation i.e. differently perturbed C-H by H-bond donation to the surfaces. The above mentioned cases are those contributing to the most intense bands of the experimental spectrum and, unsurprisingly, they represent the most stable surfaces of forsterite, *i.e.* those having the largest surface areas in crystalline nanoparticles (see Figure 5). This means that our simulated IR features of the adsorption complexes, joint which the calculation of surface energies of the bare material, can provide important hints on the morphology obtained during the crystallization of forsterite.

The broad band at 3500 cm$^{-1}$ can be ascribed to interacting silanols, originated from the deprotonation of HCN. This can happen on defective and/or reactive sites, such as surface-exposed $Mg^{2+}$–$O^{2-}$ ion pairs. From the computational standpoint, this situation can be somehow simulated by high energy forsterite surfaces, *i.e.* (111) and (021), where indeed the HCN deprotonation easily occurs, generating cases g and h ($c_i$ = 0.7301 and 0.2083). However, we cannot exclude that strong interactions between silanols and deprotonated $CN^-$ anions can produce very large OH bathochromic shift that fall into the CH stretching zone (from 3200 $cm^{-1}$ downwards), as shown by the case *f* ($\nu$(C≡N) = 3056 $cm^{-1}$, $c_i$ = 1.0000).

As regards the $CN^-$ anion, only case *i* has an appreciable weight in the fitted spectrum ($\nu$(C≡N) = 2151 $cm^{-1}$, $c_i$ = 0.2016), while all the others contributions (from *j* to *n*) are around 1% or less. In this case the $CN^-$ is bridged between two $Mg^{2+}$ cations forming a $\eta^2$(C,N) adsorption mode (Mg–C≡N–Mg) resulting in a hypsochromic shift of the CN bond, which explains the experimental band at 2137 $cm^{-1}$.

Table 1: Summary of all the contributions to the simulated IR spectrum after the best fit to the experimental one, grouped depending on the adsorption type, and ordered according to decreasing weight in the fitted linear combination. For each item we report the corresponding type of adsorption (molecular (M) or dissociative (D)), the involved stretching modes, the corresponding surface, the normalized weight in the linear combination ($c_i$) and the computed stretching frequency (in $cm^{-1}$.)

| Label | Type | Mode | Surf. | $c_i$ | $\nu_{CH}$ | $\nu_{CN}$ | $\nu_{OH}$ |
|---|---|---|---|---|---|---|---|
| a | M | CN + CH | (120) | 0.5408 | 3305 | 2099 | |
| b | M | CN + CH | (120) | 0.2173 | 3276 | 2099 | |
| c | M | CN + CH | (010) | 0.1688 | 3195 | 2064 | |
| e | M | CN + CH | (111) | 0.0074 | 2942 | 1526 | |
| f | D | OH | (021) | 1.0000 | | | 3056 |
| g | D | OH | (021) | 0.7301 | | | 3466 |
| h | D | OH | (111) | 0.2083 | | | 3571 |
| i | D | CN | (001) | 0.1965 | | 2151 | |
| j | D | CN | (021) | 0.0093 | | 2076 | |
| k | D | CN | (111) | 0.0100 | | 1996 | |

Finally, it is worth mentioning case *e* ($\nu$(C=N) = 1526 $cm^{-1}$), which, even if it does not present a good fit with the experimental IR ($c_i$ = 0.0074), we believe represents an interesting case as well, possibly explaining the band at 1600-1700 $cm^{-1}$. This is the region of the double bond stretching of C=N, among other moieties, and was assigned to a bridged species of non-deprotonated HCN where the bond order of C≡N decreases and a new covalent bond is formed between the C atom of HCN and one of the O atoms of the surface. As for the HCN deprotonation, this particular chemisorption can occur on reactive/defective sites, *i.e.* surface-exposed $Mg^{2+}$–$O^{2-}$ ion pairs, and high energy surfaces.

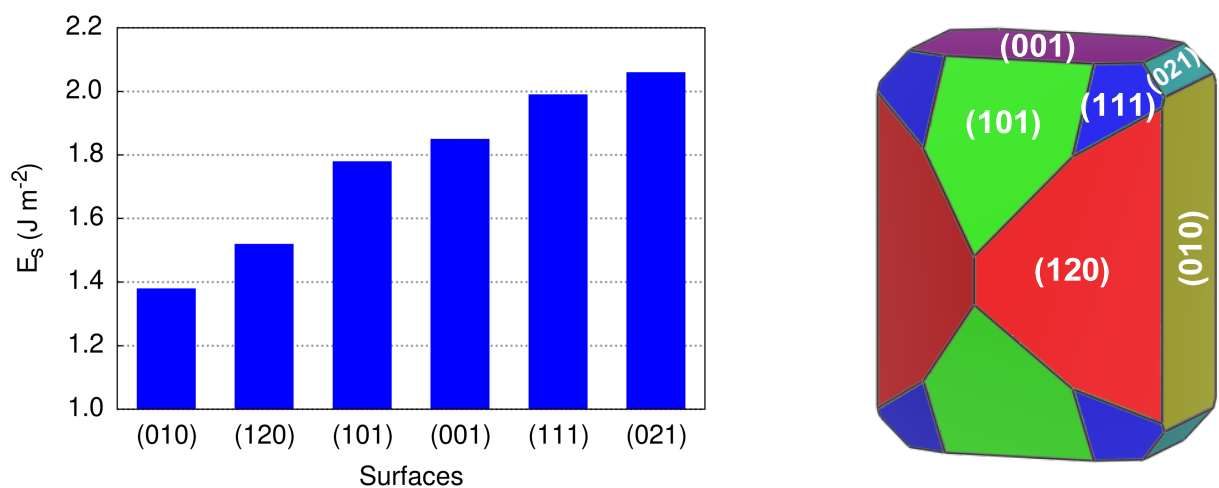

Figure 5: Left: surface energy of the six slab models adopted in this work calculated at the PBE-D*N level. Right: Wulff's construction of a forsterite nano-crystal at 0 K. Adapted with permission from 35. Copyright 2023 Royal Society of Chemistry.

## Conclusions

This combined experimental and computational study provides a comprehensive understanding of the adsorption behavior of hydrogen cyanide (HCN) on crystalline forsterite ($Mg_2SiO_4$) silicate surfaces. Through infrared (IR) spectroscopic measurements and periodic density functional theory (DFT) simulations, the intricate HCN/forsterite interactions occurring during the HCN adsorption on the crystalline forsterite surfaces at low temperatures (100–150 K) were elucidated at an atomistic detail.

The main achievement of this synergy is twofold: on one hand, we remove many biases related to the deconvolution of a complex spectrum by making a weighted convolution of all the simulated IRs, fitted on the experimental spectrum. On the other hand, the exploration of all forsterite surface models allowed us to gauge the role of each specific crystal face in contributing to the vibrational features of the experimental spectrum which, in turn, allows to elucidate the crystal morphology of the forsterite sample.

Results reveal the prevalence of both molecular and dissociative adsorption complexes, the former dominating on stable surfaces, while the latter occurring predominantly on less stable surfaces. HCN dissociation occurs due to the enhanced Lewis acid-base behaviour of surface-exposed $Mg^{2+}$–$O^{2-}$ ion pairs, but also to reactive silicate groups belonging to high energy surfaces, as it has been demonstrated by DFT simulations, which therefore present a non-negligible contribution to the exposed facets of experimental forsterite samples.

These findings are useful to deepen our understanding on the interaction of HCN with silicate surfaces and, by extension, to the catalytic properties of silicates towards HCN activation, thereby contributing to improve our know-how of prebiotic chemistry in astrophysical environments. Our combined experimental-theoretical approach can also be used to explore the adsorption properties of other molecules on different inorganic material surfaces, thereby further advancing our knowledge on surface chemical processes such as adsorption and heterogeneous catalysis relevant to fundamental (like astrochemistry and prebiotic chemistry) and more applied fields.

# Acknowledgement

This project has received funding within the European Union's Horizon 2020 research and innovation program from the European Research Council (ERC) for the project "Quantum Chemistry on Interstellar Grains" (Quantumgrain), grant agreement No. 865657. The Italian Space Agency for co-funding the Life in Space Project (ASI N. 2019-3-U.O), the Italian MUR (PRIN 2020, Astrochemistry beyond the second period elements, Prot. 2020AFB3FX) are also acknowledged for financial support. This research is also funded by the Spanish

MICINN (projects PID2021-126427NB-I00 and CNS2023-144902). This research acknowledges support from the Project CH4.0 under the MIUR program "Dipartimento di Eccellenza 2023-2027". The authors thankfully acknowledge RES resources provided by UMA for the use of Picasso (activity QHS-2023-3-0032) and by IAC for the use of LaPalma (activity QHS-2023-1-0020). The supercomputational facilities provided by CSUC and CINECA (ISCRAB projects) are also acknowledged. The EuroHPC Joint Undertaking through the Regular Access call project no. 2023R01-112, hosted by the Ministry of Education, Youth and Sports of the Czech Republic through the e-INFRA CZ (ID: 90254) is also acknowledged.

## Supporting Information Available

Electronic Supplementary Information (ESI) available: graphical representation of the forsterite slab models adopted in this work, IR spectrum with the isolated contributions to the fitting procedure and table of all the structures contributing to the fitting.